# Project Monitomation Version 1

A Novel Design For Smart Wireless Monitoring, Messaging & Task Automation Network.

Vibhutesh Kumar Singh  
Department Of ECE, IIIT Delhi  
vibhutesh.k.singh@ieee.org

Sanjeev Baghoriya  
Department Of ECE, IIIT Delhi  
sanjeev13161@iiitd.ac.in

Vivek Ashok Bohara  
Assistant Professor, IIIT Delhi  
vivek.b@iiitd.ac.in

*Abstract*— Project Monitomation is a unique implementation which focuses on justifying the capability of smart wireless networks based on IEEE 802.15.4 standard, for low power, short range Personal Area Network (PAN) communication. Through this project wireless text messaging, device control & network monitoring is implemented to demonstrate the future of Internet of things.

## I. INTRODUCTION

In this project we have implemented first of its kind Wireless Personal Area Network, which includes a text messaging system combined with multi-task automator end devices. The project presents a low-cost, power efficient, portable, secure & flexible solution for the same. It's global in the sense, that it could be easily integrated for Home, Office, Factory & Network Monitoring environment. The project consists of software front end which prepares & transmits instructions to a set of portable end devices to perform their assigned task. This set includes a specific end device, which is assigned the part of message display, apart from that; it will also do the network monitoring task. Here we have provided the flexibility to the user upon how he wants to provide input for automation tasks, as he can either use his mobile phone for that or simply type that instruction in the software. Regardless on how the input was given; the software front end will decode it to an instruction set the same way. For text messages the user has to use our software front end.

Our project also implements a secure wireless network so that any external device outside the network without knowing the network SSID will not able to invoke any of the function of end devices maliciously. Moreover, even if someone deliberately breaks in the system, the message display unit, which we are also using it for network monitoring purpose will reveal that activity. Thus this project can also be used by security agencies to intercommunicate within a small distance securely. The wireless network is implemented through embedded IEEE802.15.4 standard device [3], which are known for their sparse energy consumption [5], thus providing an energy efficient solution for above mentioned tasks. The software for our project is developed using NI LabVIEW [1]. Using it, gives developer an ease to update & build custom made automation task performing network very easily for home-office-network monitoring tasks.

## II. IEEE 802.15.4 PROTOCOL OVERVIEW

The IEEE802.15.4 standard relies on two services that Physical Layer provides. The Physical Layer Data service & management service. The Physical Layer Data service controls the transmission and reception part of the physical layer data units whereas management service perform Energy Detection tasks, carrier sense before transmitting messages and quality indication for the received packets[8]. The original version of this standard specifies two physical layers based on direct sequence spread spectrum (DSSS) one of them works in the frequency band of 868/915 MHz & baud rate of 20 to 40 Kbit/s, other in frequency band of 2450 MHz with a baud rate of 250 Kbit/s[8]. We have used 2450 MHz band for our project.

The MAC layer in IEEE 802.15.4 standard provides interface between the Service Specific Convergence Sublayer and the physical layer[8] & relies on two services, MAC data service which is responsible for the transmission and reception of the MAC physical data layer units via physical layer data service & the MAC management service, which manages the network beacons. MAC layer is also responsible for personal area network connection and disconnection, frame validation, and providing acknowledgment signal. It uses Carrier Sense Multiple Access with Collision Avoidance technique for channel access. MAC layer also applies and maintains the Guaranteed Time Slot mechanism and supports device security.

| Frequency Band (MHz) | Spreading Parameters | | Data Parameters | | |
|---|---|---|---|---|---|
| | Chip rate (Kchips/s) | Modulation | Bit Rate (kb/s) | Symbol rate (Sym/s) | Symbols |
| 868-868.6 | 300 | BPSK | 20 | 20k | Binary |
| 902-928 | 600 | BPSK | 40 | 40k | Binary |
| 2400-2483.5 | 2000 | O-QPSK | 250 | 62.5k | 16-ary Orthogonal |

**Table.1** Frequency Band & Modulation scheme our hardware can support [4]

## III. IMPLEMENTATION

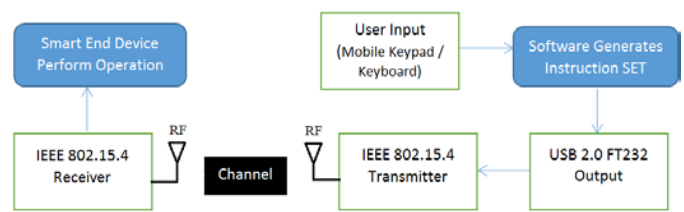

Figure 1. Schematic of Our Implementation

### A. Hardware & Software

For the software to be installed, it needs NI LabVIEW runtime installed which has a minimum requirement of any Windows, MAC or Linux based PC, having at least a Keyboard & sound card installed [2]. Also it must meet the following minimum hardware requirement: 866 MHz processor, 256 Mb RAM, 1024x768 pixels screen resolution, & 500Mb Disk Space [2]. As in addition to keyboard our software also supports Dual-Tone Multi-Frequency (DTMF) tone input, there is a need for at least a Microphone or double headed TRS connector for that feature to be used. We would recommend the use of double headed TRS connector wires as it would provide a highly accurate result due to low channel noise. The software sends the information through USB FT232 ports for which the BAUD rate can vary from 2400 – 115200 bps [7]. From USB end the instruction is then sent to IEEE802.15.4 Transmitter & is received by IEEE802.15.4 Receiver, & the device performs the intended function, as shown in Figure 1. We have used Atmega2560 & Atmera328 microcontrollers to develop our smart end device.

### B. Software & its algorithm

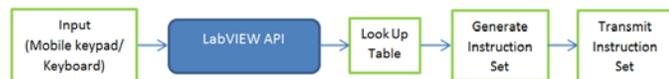

**Figure 2**. Software Instruction Flow

Our software is responsible for interaction with the user. In this we have provided the option to the user to either provide the input through the keypad of any mobile phone or by typing the instruction from the keyboard. After getting the input from the user it will traverse through a look-up table to find whether it's a special controlling input, if not it will send that input as a text otherwise will parse it in form of a standard instruction & transmit, refer Figure 2. Additionally we need at least NI LabVIEW 7.0 run time, & USB FT232 drivers-preinstalled installed. The end devices are dynamically programmed to gather all the transmitted data in the channel, however to act only when it is instructed to do so otherwise discard the gathered data, thus making them smart.

The main benefit of using NI-LabVIEW Development Environment is the software patching/updating part. Thus developing user defined custom applications also tend to become a lot easier as compared to other current SDKs available, due to its modular nature .

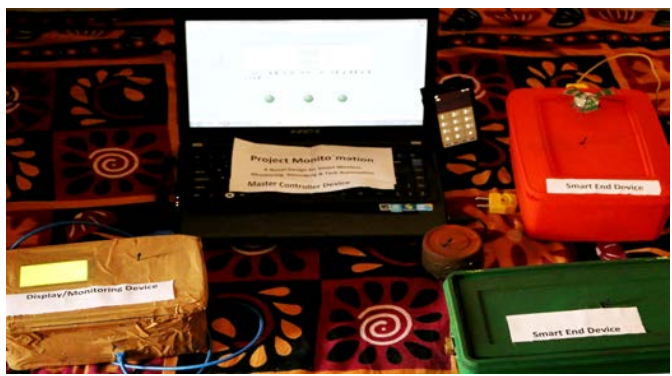

Figure 3. Our Implementation (boxed), showing Controlling & Smart End Devices at their root level

### IV. MOTIVATION & PROBLEM SOLVING

Lot of home, office, factory as well as other commercially available appliances uses some sort of remote control device to execute a instruction from the user, however major drawback of such a device is lack of security, inability to support large number of nodes and durability. Furthermore, neither these devices are able to transmit text messages & monitor what's going on, in the network. use some remote control device & the main problems with these devices is the lack of security, inability to support large number of nodes & longevity [6], where as in our system the number of nodes supported can go up to 256 including coordinator device [5] , secure & low power. Neither these implementation are able to transmit text messages & monitor what's going on, in the network. Hence the motivation behind this work is to design a cost effective, low power and secure wireless network [5] which could be controlled by a PC, a mobile phone or both, so that one can easily use in home-office environment. Additionally it should have a text based support and can support large number of nodes (upto 256).

Eventually this kind of design can be a good head-start for the Internet of Things (IOT). For example ,a person is sitting in bed room can monitor the status of lights, fans, doors, music player, home appliances & can control them right from his mobile phone or through our software. Our design for such wireless network provides a cost effective solution.

### V. APPLICATIONS

Project Monitomation can do multitasking as in it can do text messaging, device control, network monitoring, data transmit & automation of innumerous functions in a single implementation. Our project is best suited for the home-office-factory ambience as it can inform & alert the users with a message, & also if there is any malicious activity happening in their home/office network. It can track tagged assets to the exact location with an ability of duplex communication. It can very well be implemented in setting up smart meters, smart traffic management systems & wherever the short range communication find a valuable use.